\documentclass[conference]{IEEEtran}
\IEEEoverridecommandlockouts

\usepackage{cite}
\usepackage{amsmath,amssymb,amsfonts}
\usepackage{algorithm}
\usepackage{algorithmic}
\usepackage{graphicx}
\usepackage[caption=false,font=footnotesize]{subfig}
\usepackage{booktabs}
\usepackage{array,tabularx}
\usepackage{multirow}
\usepackage{textcomp}
\usepackage{framed}
\usepackage{xcolor}
\usepackage{xspace}
\usepackage{url}
\usepackage[hidelinks]{hyperref}

\usepackage{pifont}
\usepackage{wasysym}
\usepackage{xcolor}

\newcommand{\sYes}{{\ding{51}}}
\newcommand{\sHalf}{\LEFTcircle}
\newcommand{\sNo}{{\ding{55}}}
\DeclareRobustCommand{\NetConfArena}{NetConf\-Arena\xspace}

\newcommand{\nn}{\textbackslash{}n\allowbreak}

\def\BibTeX{{\rm B\kern-.05em{\sc i\kern-.025em b}\kern-.08em
    T\kern-.1667em\lower.7ex\hbox{E}\kern-.125emX}}
\begin{document}
\raggedbottom

\title{\NetConfArena: An Executable Benchmark for LLM Agents in Closed-Loop Network Configuration}

\author{
\IEEEauthorblockN{Chang Liu, Xiaohui Xie\textsuperscript{*}, Xinyi Chen, and Yong Cui\textsuperscript{*}}
\IEEEauthorblockA{Department of Computer Science and Technology, Tsinghua University, Beijing, China}
\thanks{This work was supported by NSFC Project under Grant 62132009 and Grant 62221003.}
\thanks{\textsuperscript{*}Corresponding Authors: Xiaohui Xie and Yong Cui.}
}

\maketitle

\begin{abstract}
Large language model (LLM) agents are increasingly attractive for automating network configuration, yet their reliability and failure patterns are poorly understood.
An essential prerequisite is to assess such agents in a realistic but risk-free environment.
Existing benchmarks, however, fall short: they often treat configuration as static command generation or rely on overly simplified settings.
Such evaluations understate the core challenges of network configuration, where correctness requires reasoning about protocol complexity and topology dependence.
We present \NetConfArena, an executable benchmark for evaluating LLM agents in closed-loop network configuration.
\NetConfArena places agents in emulated multi-device networks, provides a standardized and compact action interface for task execution, and evaluates the resulting network behavior with hidden task-specific executable test cases.
The benchmark relies on an LLM-assisted, emulation-grounded pipeline, which converts human-oriented network materials into reusable parameterized task templates.
We evaluate representative LLM agents on 480 task instances instantiated from 96 protocol-focused task templates, yielding 3840 execution trajectories, and show that failures are not limited to command errors.
The failures also reveal gaps in task-specification adherence and robust planning and execution.
These findings suggest two future directions: using validated trajectories as supervision signals to improve foundation models, and designing harness mechanisms that make agent execution more reliable and accountable.
\end{abstract}

\section{Introduction}

Network configuration is a core operational task for maintaining connectivity, policy compliance, and predictable routing.
Although operators often express their intent at a high level, realizing it usually requires coordinated updates across multiple devices and protocols~\cite{fogel2015general,beckett2017general,el2017network,el2018netcomplete,beckett2016don,beckett2017network,tian2019safely,ramanathan2023practical}.
Therefore, configuration correctness cannot be equated with whether configuration commands are accepted by a single device; it depends on whether the resulting network behavior satisfies that intent.

Recent progress in large language models (LLMs) has created substantial interest in using LLM agents to automate network configuration~\cite{long2025survey,liu2024large}.
Such agents appear well suited to this setting because they can interpret natural-language intent, reason over intermediate observations, and invoke tools over multiple steps~\cite{yao2023react,openai2024learning}.
In practice, however, few operators would be willing to let an LLM agent configure a production network directly.
This caution is warranted: network configuration is error-sensitive, and an unintended change can disrupt network-wide connectivity and cause service outages.
Without a trustworthy understanding of LLM agent reliability and failure modes, direct deployment in such a high-stakes setting remains difficult to justify.
Accordingly, a necessary first step is to evaluate these agents in a risk-free environment~\cite{yao2023react,yang2024swe,lai2024autowebglm,zhang2025webpilot,hong2023metagpt,liu2025pc}.
Such assessment must also reflect how network configuration is actually performed.
The agent should not be judged only by its final output, because network configuration is naturally an iterative process in which operators observe the current network state, apply changes cautiously, check their effects carefully, and correct mistakes promptly when the network does not behave as expected.
Moreover, network configuration does not have a unique valid solution, as many different command sequences can be acceptable if they lead to the intended network state.
For these reasons, evaluating LLM agents for network configuration requires closed-loop interaction and behavioral validation, rather than static comparison with a single reference configuration.

However, existing benchmarks for LLM-based network configuration still fall short of capturing realistic closed-loop configuration in three key respects.
First, benchmarks based on synthetic or manually labeled datasets typically evaluate static outputs against reference answers, primarily testing whether a model has covered the required configuration knowledge~\cite{wang2024netconfeval}.
Second, benchmarks that execute generated configurations in emulators often use emulation only for post hoc validation, thereby measuring intent understanding and single-shot generation more than interactive execution, diagnosis, and repair~\cite{aykurt2024netllmbench}.
Third, the closest closed-loop settings are often limited to simplified single-router scenarios, which do not expose the protocol complexity or topology dependence of realistic network configuration~\cite{zhou2026netarena}.
Consequently, existing evaluations provide limited evidence on whether current agents can reliably complete network configuration through vendor-specific command-line interfaces (CLIs).

To address this gap, we present \NetConfArena, an executable benchmark for evaluating LLM agents in closed-loop network configuration.
Compared with prior benchmarks, \NetConfArena goes beyond simplified settings toward more realistic scenarios by supporting a vendor-specific CLI configuration process and multi-device, protocol-rich tasks.
\NetConfArena relies on an LLM-assisted, emulation-grounded pipeline to build a task suite that is executable and reusable.
The pipeline starts from human-oriented network materials, including vendor manuals, public tutorials, and community lab scenarios, and converts them into validated and parameterized templates, allowing controlled variations of addresses, interfaces, and protocol parameters without changing the underlying intent or test-case logic.
During evaluation, \NetConfArena places agents in emulated networks and gives them a standardized and compact action interface for task execution.
Each task specifies an operator intent together with an executable environment context, and the submitted result is evaluated by hidden task-specific test cases that check the final network behavior.
In addition to outcome scores, \NetConfArena records complete interaction trajectories, making it possible to analyze how an agent reaches the final state.

Using this benchmark, we evaluate representative LLM agents on 480 seeded task instances derived from 96 task templates and collect 3840 execution trajectories.
The results show that agents built on state-of-the-art foundation models can solve many tasks, but their reliability varies across different settings.
For agents built on less capable models, failures are not limited to misunderstandings of protocol or command semantics.
Instead, many failures arise because agents deviate from task-provided specifications, plan and execute ineffectively, or give up after unexpected feedback.
These findings show that realistic network configuration stresses not only command generation, but also specification adherence and robust closed-loop execution.

The trajectories produced by \NetConfArena also point to future directions beyond benchmarking.
Successful trajectories can serve as validated supervision signals for improving foundation models, while failed trajectories and process metrics can guide the design of better agent harnesses.
In particular, harness mechanisms that improve state management and enforce safety constraints may make agent execution more reliable and accountable.
In this sense, \NetConfArena is not only a measurement tool, but also a source of high-quality validated trajectories for model training and practical guidance for agent-system design.
We have released the \NetConfArena benchmark framework and task suite as open source\footnote{\url{https://github.com/liujona/NetConfArena/}.} to support reproducible evaluation and future research.

In summary, this paper makes the following contributions:
\begin{itemize}
    \item We develop an emulator-backed benchmark framework in which agents try to complete configuration tasks with vendor-specific CLIs.
    The framework also combines a standardized and compact action interface with hidden executable test cases to support evaluation from both process- and outcome-level perspectives (Sec.~\ref{sec:netconfarena}).

    \item We design an LLM-assisted, emulation-grounded pipeline that transforms human-oriented materials into validated and parameterized task templates.
    Using this pipeline, we construct 96 protocol-focused task templates with controlled variation and deterministic test cases (Sec.~\ref{sec:benchmark_task}).

    \item We evaluate representative LLM agents over 3840 execution trajectories, identify four major knowledge-level and interaction-level failure patterns, and discuss two directions for improving agent reliability (Secs.~\ref{sec:evaluation} and~\ref{sec:future_directions}).
\end{itemize}

\section{Background and Motivation}
\label{sec:motivation}

\subsection{Motivation Example}
Network configuration is protocol-aware and topology-dependent, and its outcomes are sensitive to the procedural steps taken.
% This property calls for closed-loop, behavior-grounded evaluation rather than static output comparison.
We illustrate this property with a routing-manipulation task from \NetConfArena and its corresponding successful execution trace.

\begin{figure}[t]
  \centering
  \includegraphics[width=\linewidth]{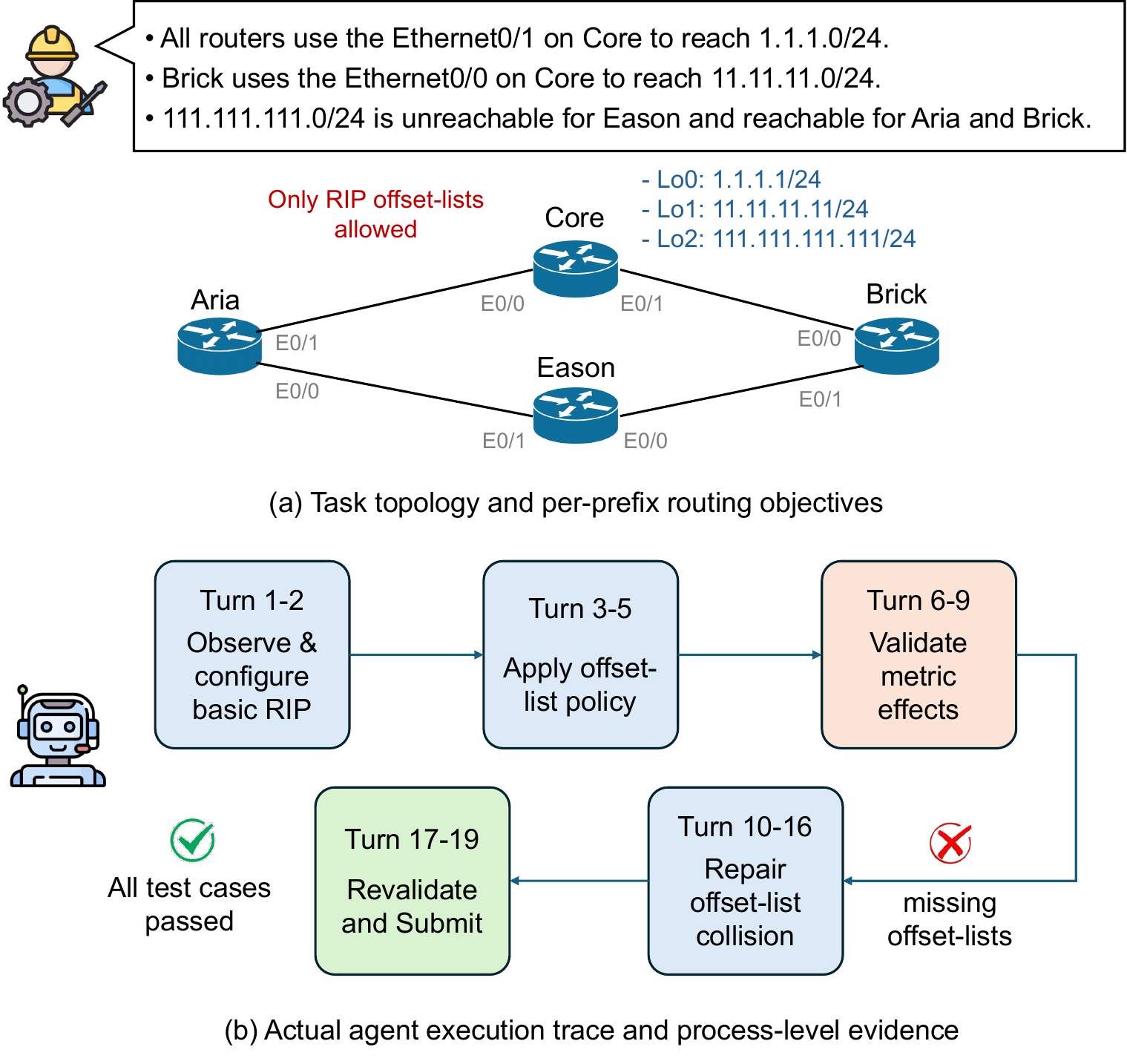}
  \caption{Motivation example from \NetConfArena, illustrating the closed-loop workflow of a network configuration agent.}
  \label{fig:motivating}
\end{figure}

Fig.~\ref{fig:motivating} (a) presents the task topology and its fine-grained, per-prefix routing intent.
Although the intent is concise, satisfying it requires coordinated configuration of several interdependent elements, including ACLs that match the target prefixes and RIP offset-lists that correctly reference those ACLs.
It also requires deciding where each offset-list should be placed, including the router, interface, and whether it should be applied to incoming or outgoing RIP updates.
The metric adjustment must then be carefully chosen so that the resulting route selection and reachability satisfy the intent.
These elements are tightly coupled: an error in any one can make the final network behavior deviate from the objective.

To complete such a task, experienced network engineers typically decompose the workflow into progressive configuration, intermediate validation, and feedback-driven repair.
Fig.~\ref{fig:motivating} (b) illustrates a successful closed-loop execution that follows this pattern: the agent observes and reasons about the network state, applies protocol-aware changes, checks whether each applied action has taken effect, and uses feedback to refine the configuration until all routing constraints are satisfied.
Some feedback is available directly from command execution, such as rejected commands.
Other feedback requires active validation, in which the agent issues diagnostic commands to verify configuration and network-level effects such as ACL matching, metric changes, and selected next hops.
Rather than stopping after generating an initial configuration, an intelligent agent should use this feedback to refine its decisions and progress toward the final network objective.

This example motivates evaluating both the final network behavior and the process that produced it.
The final outcome shows whether the submitted configuration satisfies the operator intent.
The process shows whether the agent inspected the right state, used feedback from the environment, and repaired observed mistakes.
% , and reached the successful state through grounded interaction rather than by producing a plausible but unchecked command sequence.

\subsection{Task Formulation}
\label{sec:task_formulation}

We formulate network configuration as a closed-loop decision-making task between an LLM agent and an executable network environment~\cite{yao2023react,zhou2026netarena}.
A task in \NetConfArena is defined as
\[
\tau = \langle I, E, A, C \rangle ,
\]
where \(I\) denotes the operator intent, \(E\) denotes the network environment, \(A\) is the set of actions available to the agent, and \(C\) is a set of hidden executable test cases used for evaluation.

The intent \(I\) is expressed in natural language and may contain multiple sub-intents, such as enforcing reachability, selecting a specific routing path, or controlling route propagation.
The environment \(E\) specifies the network topology, the initial device configurations, and the vendor-specific command syntax and protocol semantics that constrain the agent's actions.
The agent interacts with the environment through the action space \(A\) and receives an observation after each step.
These actions support information gathering, configuration, validation, and task submission.
We define the specific actions and their corresponding observations in Sec.~\ref{sec:interactive_execution}.
% The resulting interaction is closed-loop: the agent can revise its subsequent actions according to the observed network state and device feedback rather than relying only on the initial task description.

An execution of an agent on task \(\tau\) produces a task-specific trajectory
\[
\xi_\tau = (o_0, r_0, a_0, o_1, r_1, a_1, \ldots, o_T, r_T, a_T),
\]
where \(o_t\) is the observation returned by the environment at step \(t\), \(r_t\) is the agent-generated rationale or plan, and \(a_t \in A\) is the action issued to the environment.
The trajectory ends when the agent issues \texttt{submit}, or when the environment terminates execution after the interaction limit is reached.

It is worth emphasizing that agents make decisions based on observations rather than on a global state representation, so a formal state definition is not part of the agent's decision loop.
The final environment state instead serves as the object of evaluation.
Once the agent execution ends, \NetConfArena executes the intent-specific test cases in \(C\) to examine whether the final network behavior satisfies the original intent \(I\).
It also considers process-level signals from the trajectory, such as whether the agent issues invalid actions, performs informative validation, and repairs observed errors before submission.

\subsection{Existing Benchmarks Are Insufficient}
Based on the discussion above, a holistic evaluation framework for LLM agents on network configuration tasks should satisfy three core requirements.
\begin{itemize}
    \item \textbf{R1: Closed-loop interactive environment.}
    Agents should be able to observe network state, apply configurations, validate their effects, and iteratively repair failures based on feedback.
    \item \textbf{R2: Realistic configuration scenarios.}
    Tasks should require protocol-aware reasoning, vendor-specific CLI usage, and coordination across multiple devices.
    \item \textbf{R3: Outcome and process evaluation.}
    Evaluation should assess final network behavior and whether the agent follows a reasonable, safe, and goal-directed execution process.
\end{itemize}

% ===== 表格本体 =====
\begin{table}[t]
  \centering
  \caption{Comparison with existing benchmarks for LLM-based network configuration.}
  \label{tab:benchmark_comparison}
  \small
  \setlength{\tabcolsep}{16pt}
  \begin{tabular}{lccc}
    \toprule
    Benchmark & R1 & R2 & R3 \\
    \midrule
    NetConfEval & \sNo & \sHalf & \sNo \\
    NetLLMBench & \sNo & \sNo & \sHalf \\
    NetArena & \sYes & \sNo & \sHalf \\
    \NetConfArena & \sYes & \sYes & \sYes \\
    \bottomrule
  \end{tabular}
  \vspace{2pt}
  
  \footnotesize{\sNo~= not supported, \sHalf~= partially supported, \sYes~= supported.}
\end{table}

As summarized in Table~\ref{tab:benchmark_comparison}, existing benchmarks address different aspects of LLM-based network configuration, but none jointly satisfies R1--R3 in an integrated setting for realistic, closed-loop evaluation of network configuration agents.
NetConfEval~\cite{wang2024netconfeval} examines whether LLMs can assist network configuration through tasks such as specification translation, API generation, and low-level command synthesis.
Nevertheless, its evaluation remains largely static: models are mainly assessed on manually generated datasets rather than through iterative interaction with an evolving network state.
NetLLMBench~\cite{aykurt2024netllmbench} advances executable evaluation by validating LLM-generated configurations in emulator-based environments, rather than relying solely on textual comparison.
Yet the emulator mainly serves as a validation backend rather than a closed-loop interactive environment.
NetArena~\cite{zhou2026netarena} is the closest to our setting, as it studies LLM agents in closed-loop network environments.
However, its scenarios remain substantially abstracted.
Its tasks run on single-router topologies with a few flat subnets, and configuration operations are performed through Linux system commands such as \texttt{sysctl}, \texttt{iptables}, and \texttt{ip route}, rather than vendor-specific CLIs used in production networks.

Taken together, these limitations leave a critical gap: existing benchmarks do not adequately test whether LLM agents can handle protocol-aware, multi-device configuration tasks under evolving network states.
% This gap motivates the design of \NetConfArena as an interactive benchmark that evaluates LLM agents in emulated network environments at both the outcome and process levels.

\section{\NetConfArena Framework}
\label{sec:netconfarena}
% This section details the design and workflow of \NetConfArena.
% We first introduce its overall architecture and the design choices behind it, then describe how it sets up executable tasks, supports closed-loop interaction through a standardized and compact action interface, and evaluates agents from both outcome-level and process-level perspectives.

\subsection{Design Overview}

To operationalize the three requirements established in Section~\ref{sec:motivation}, \NetConfArena is built around three design choices.
First, agent execution and evaluation are performed in high-fidelity emulated networks, where agents configure multiple devices using vendor-specific CLIs.
This setting exposes agents to the command syntax and cross-device protocol dependencies that realistic configuration tasks require.
Second, the outcome of each run is determined by hidden deterministic test cases.
This yields objective and reproducible scoring, avoiding the subjectivity of LLM-based judging and the scalability limitations of human evaluation.
Third, the agent action space is deliberately compact, exposing only five abstract actions.
Beyond keeping the benchmark neutral with respect to agent architecture, the compact action space makes failures easier to attribute, because the role of each action type is well defined.

\begin{figure}[t]
  \centering
  \includegraphics[width=\linewidth]{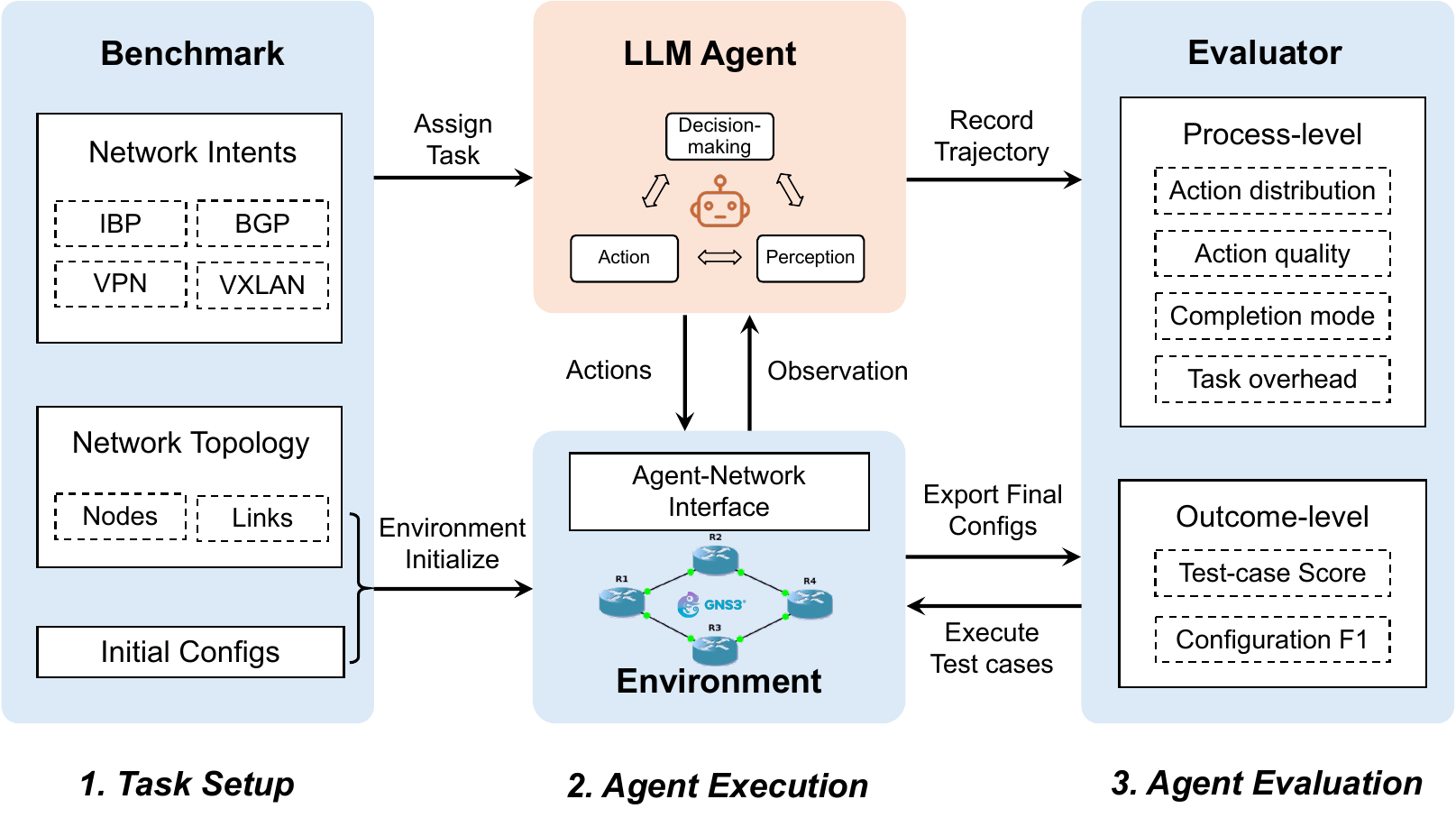}
  \caption{
  Overview of the \NetConfArena framework.
  \NetConfArena connects the Benchmark, LLM Agent, Environment, and Evaluator components across three stages: task setup, agent execution, and agent evaluation.
  }
  \label{fig:overview}
\end{figure}

Figure~\ref{fig:overview} presents the architecture of \NetConfArena, which comprises four major components that interact across three stages: task setup, agent execution, and agent evaluation.
\begin{itemize}
    \item \textbf{Benchmark:} a set of parameterized task templates, each of which defines network intents, topologies, initial device configurations, and task-specific test cases.
    \item \textbf{LLM Agent:} the evaluation target\footnote{\NetConfArena does not prescribe a specific agent architecture.
    In our experiments, we instantiate this component with a standard ReAct-style agent~\cite{yao2023react}.} that receives the task intent and interacts with the environment through standardized actions for perception, configuration, validation, and submission.
    \item \textbf{Environment:} a task-specific network environment that instantiates the topology in an emulator, loads the initial configurations, and responds to agent actions by updating the network state and returning observations.
    \item \textbf{Evaluator:} a module that computes outcome-level metrics by executing task-specific test cases, and process-level metrics by analyzing the task execution trajectory.
\end{itemize}

\subsection{Task Setup}

For each evaluation run, \NetConfArena instantiates a parameterized task template according to the parameterization rules defined in Section~\ref{sec:benchmark_task}.
This process produces a concrete task instance, including the operator intent, topology specification, initial device configurations, reference configuration, and task-specific test cases.

\NetConfArena separates the information exposed to the agent from the information reserved for evaluation.
During execution, the agent receives the exposed task context, including the operator intent and topology information, and can inspect or modify the network through the standardized action interface.
The reference configuration and test-case definitions remain hidden from the agent.
These reserved artifacts are used after submission to evaluate the final network state in a consistent and repeatable manner.

For each instantiated task, the setup stage provides all agents with the same executable environment and initial network state.
Across task instances, \NetConfArena can vary addresses, interfaces, devices, or protocol parameters in a controlled manner while preserving the underlying operator intent and test-case logic.
This design tests whether an agent can solve the configuration problem in the running network, rather than memorizing a fixed textual configuration.

\subsection{Interactive Agent Execution}
\label{sec:interactive_execution}
% Interactive agent execution is the core stage of the \NetConfArena workflow.
In this stage, an agent interacts with a running emulated network through a compact action interface, observes realistic device feedback, and makes sequential decisions based on the evolving network state.

\subsubsection{Emulator-backed execution environment}
\NetConfArena provides an emulator-backed execution environment for each task.
In our implementation, this component is built on GNS3, an open-source network emulation platform that allows us to programmatically create isolated network topologies for individual tasks~\cite{gns3}.
By running vendor router images in GNS3, \NetConfArena emulates vendor-specific command syntax, device feedback, and protocol dynamics.
This setup supports agents in carrying out realistic configuration workflows while keeping evaluations controlled and reproducible.

\subsubsection{Agent-network interface}
Following the workflow of network operators, \NetConfArena provides an agent-network interface consisting of five abstract actions: \texttt{get\_running\_config}, \texttt{apply\_config}, \texttt{execute\_validation}, \texttt{wait}, and \texttt{submit}.
% We design this action set to be compact and semantically distinct, yet sufficient for end-to-end configuration tasks.
Except for \texttt{submit}, which ends the interaction and triggers task evaluation, every action returns an observation that the agent uses to ground its next decision.

\texttt{get\_running\_config} retrieves the current configurations of selected devices and returns the full configuration text of each device.
\texttt{apply\_config} applies device-level configuration commands and returns the device feedback for each command, including acceptance or rejection messages.
It automatically enters and leaves configuration mode on each target device, so the agent only supplies the configuration commands themselves and receives the resulting command-by-command echo.
\texttt{execute\_validation} runs non-configuration diagnostic commands such as \texttt{show ip route} and \texttt{ping}, and returns their raw device output.
Such output exposes the protocol state the agent needs in order to judge its progress, including routing and neighbor tables, interface status, and end-to-end reachability.
\texttt{execute\_validation} is available to the agent during interaction and is distinct from the hidden evaluator test cases used after submission.
All of these actions can be issued to multiple devices in a single interaction turn when the agent needs coordinated inspection, configuration, or validation.
\texttt{wait} lets the agent pause for protocol convergence or delayed device state changes and returns a notification once the pause expires.

\subsubsection{Closed-loop interaction}
During task execution, the agent repeatedly observes the current network state and decides the next action.
% This closed-loop interaction is essential for evaluating whether an agent can use feedback to make effective sequential configuration decisions.
To bound the execution process and prevent agents from running indefinitely, \NetConfArena enforces a maximum number of interaction turns for each task.

During this interaction, \NetConfArena records every execution as a structured trajectory.
Each trajectory consists of the task context, the agent's reasoning, the actions it issues, and the observations returned by the environment.
This logging serves two purposes.
First, it supports failure diagnosis by exposing where an agent issues invalid commands, performs insufficient validation, or submits prematurely.
Second, it provides structured data for the agent-improvement methods discussed in Section~\ref{sec:future_directions}: successful trajectories can be used as demonstrations, while failed trajectories can be used to construct preference pairs~\cite{rafailov2023dpo}.

\subsection{Dual-Perspective Agent Evaluation}
\label{sec:dual_perspective_eval}

\NetConfArena evaluates each run from two complementary perspectives.
% The outcome perspective measures whether the final network state satisfies the task intent, while the process perspective measures how the agent reaches that state through interaction.

\subsubsection{Outcome-level evaluation}
After submission, the evaluator executes predefined task-specific test cases against the final network state.
The evaluator also computes configuration changes and compares them with the reference configuration.
\begin{itemize}
  \item \textit{Test-case score}
  This is the primary correctness metric, defined as the fraction of task-specific test cases passed by the final network state.
  It directly measures task completion in terms of the behavior induced by the executed configuration.
  A score below one indicates partial progress, and only a full score counts as passing the task.
  \item \textit{Configuration F1}
  This metric computes precision over the agent's actual configuration changes and recall over the reference configuration.
  We report this configuration comparison as a contrast metric to show that text matching alone is insufficient for judging whether the intended network behavior is realized.
\end{itemize}

\subsubsection{Process-level evaluation}
The recorded trajectory captures how the agent uses the closed-loop interface before the run terminates, providing richer evidence of its progress toward task completion.
The following process metrics reflect the efficiency and reliability of the agent's execution process.
\begin{itemize}
  \item \textit{Turn counts and action proportions}
  These metrics describe how many interaction turns a run consumes and how those turns are distributed across action types.
  We report the fraction of turns spent on configuration retrieval, configuration update, validation, waiting, submission, and null or unrecognized actions.
  \item \textit{Action quality}
  These metrics characterize whether the agent uses configuration and validation actions correctly during interaction.
  The explicit error-action rate counts parsed configuration or validation actions that are rejected by the device or environment.
  These failures often arise from limited familiarity with vendor-specific CLI command syntax and usage.
  The repeat-action rate counts consecutive turns that issue identical configuration or validation actions.
  Such repetition indicates that the agent is not converting environment feedback into new progress.
  \item \textit{Completion mode}
  This metric qualitatively distinguishes the outcome of a single task execution.
  We classify each trajectory, in order of precedence, as early submission, interaction-limit hit, correct submission, or error submission.
  Early submission denotes a \texttt{submit} action issued with no configuration attempt over the entire run, while an interaction-limit hit denotes a run that reaches the interaction limit without submitting.
  Among active submission attempts, we distinguish correct and error submissions based on whether the agent passes the task.
  \item \textit{Task overhead}
  This metric records the token overhead and elapsed wall-clock time required by an agent for each task execution.
\end{itemize}

\section{Benchmark Task Construction}
\label{sec:benchmark_task}

\begin{figure*}[!t]
  \centering
  \includegraphics[width=\linewidth]{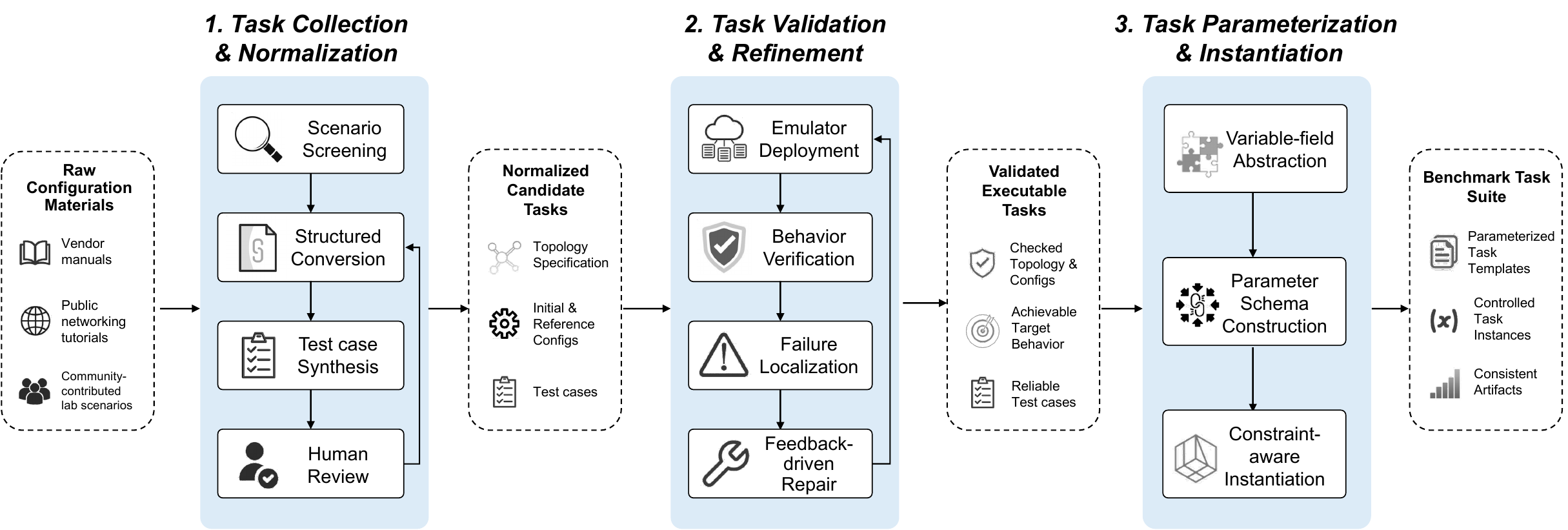}
  \caption{Benchmark task construction pipeline in \NetConfArena: human-oriented configuration materials are converted into structured task specifications with executable test cases, validated through emulation, and parameterized under constraints to produce reusable templates and seeded task instances.}
  \label{fig:benchmark_task}
\end{figure*}

% The preceding section describes how \NetConfArena organizes task setup, interactive execution, and dual-perspective evaluation.
% This section focuses on the construction of the benchmark tasks, as task quality directly affects the reliability and interpretability of agent assessment.

\subsection{Challenges and Opportunities}
Constructing a high-quality benchmark for interactive network configuration is challenging.
Each task needs not only a natural-language intent, but also a concrete topology, valid initial device configurations, and executable test cases.
Manually authoring such tasks is labor-intensive: even a small configuration objective may require carefully matching interface names, address plans, and routing processes.
% Moreover, subtle mistakes in the initial state or executable test cases can make a task unsolvable or incorrectly scored.

Fortunately, we observe that many vendor manuals, public networking tutorials, and community-contributed lab scenarios contain realistic configuration examples for common network protocols~\cite{cisco_netacad,gns3vault}.
These materials provide a valuable starting point because they reflect recurring operational intents.
However, they cannot be directly used as benchmark instances.
They are typically written for human readers, use heterogeneous formats, often describe topologies through figures rather than machine-readable specifications, and rarely provide standardized initial states or executable evaluation criteria.
Most importantly, they usually present a single worked example rather than a reusable task family, making them unsuitable for scalable and controlled evaluation.

We therefore propose an LLM-assisted, emulation-based pipeline that transforms such materials into executable, validated, and parameterized benchmark tasks.
As shown in Figure~\ref{fig:benchmark_task}, the pipeline consists of three stages, each addressing one of these limitations: task collection and normalization extracts task semantics from heterogeneous materials (Sec.~\ref{sec:task_collection}), emulation-based validation and refinement ensures the reliability of initial states and test cases (Sec.~\ref{sec:task_validation}), and constraint-guided task parameterization turns each validated task into a reusable family of instances (Sec.~\ref{sec:task_param}).

\subsection{Task Collection and Normalization}
\label{sec:task_collection}

We first use an LLM to screen raw configuration scenarios collected from vendor manuals, public networking tutorials, and community-contributed lab scenarios.
The screening step determines whether each scenario contains the information needed to define a complete benchmark task, including the topology, configuration intent, and reference configuration.

For retained scenarios, the LLM converts topologies from diagrams or informal descriptions into structured JSON specifications.
It then compares the configuration intent with the reference configuration to identify which configuration elements belong to the initial state and which encode the target behavior.
Given few-shot examples written by human experts, the LLM further decomposes the configuration intent into checkable sub-intents and derives one executable test case for each of them.
These test cases assess network behavior, such as advertised networks, selected next-hop interfaces, and route reachability.
Each of them is evaluated by executing a device-level diagnostic command on the emulated device and checking the returned output against a deterministic predicate.
Table~\ref{tab:testcase_example} in Appendix~\ref{app:template_example} shows representative test cases used for the BGP path-control template.
It is important to note that negative test cases, which check for the absence of undesired behavior, are derived only from prohibitions stated explicitly in the intent; consequently, negative-test coverage is not exhaustive.
A submitted configuration could therefore pass all test cases while introducing unchecked side effects, such as unintended route leakage or broken reachability outside the task scope.
This is an inherent limitation of evaluating network behavior against a finite set of predicates, and we plan to address it in future work (Sec.~\ref{sec:limitations}).

Human reviewers then inspect the LLM-generated task specification, focusing on whether the intent is unambiguous, the initial configuration is sufficient, and the generated test cases are aligned with the configuration intent.
Any issues they identify are provided as feedback to the LLM, which revises the task specification accordingly.
This inspection-and-revision loop repeats until the reviewers identify no further issues.

\subsection{Task Validation and Refinement}
\label{sec:task_validation}
The previous stage settles the semantics of a candidate task specification, but it cannot be considered a reliable benchmark task until it is executed and checked in a running network.
Automated construction scripts therefore deploy each candidate task in the emulator to verify the consistency of the topology and initial configurations.
The scripts then apply the reference configuration and run the task-specific test cases to confirm that the intended behavior is reachable.

This emulation-based check is especially important for test cases, which are reused across task instances and directly determine scores.
It tests their predicates against the actual behavior of devices and protocols, including whether the returned device output is correctly parsed and whether each predicate evaluates the resulting state as intended.
Grounding test cases in real device behavior reduces the risk that an LLM-generated test case preserves the same protocol misconception that an evaluated agent might have.
When validation fails, we feed the device feedback and test-case outputs back to the LLM to help localize the source of the problem.
The LLM then revises the topology specification, reference configuration, or test-case logic, and the automated scripts re-check the revised candidate task.
This loop continues until the task reliably reproduces the intended behavior in the emulated environment.

% After this stage, each task becomes a validated executable task with a well-defined initial state, an achievable target behavior, and test cases that can reliably evaluate whether the behavior has been achieved.

\begin{algorithm}[t]
\caption{Instantiating a Parameterized Task Template}
\label{alg:task-instantiation}
\begin{algorithmic}[1]
\STATE \textbf{Input:} Parameterized task template $\mathcal{T}$, value registry $\mathcal{V}$, device capabilities $\mathcal{C}$, seed $s$
\STATE \textbf{Output:} Executable task instance with consistent artifacts
\STATE $\mathcal{B} \gets \mathcal{T}.\mathrm{base}$
\STATE $\mathcal{S} \gets \mathcal{T}.\mathrm{schema}$
\STATE $\mathcal{R} \gets \emptyset$
\STATE $\mathcal{P} \gets \mathrm{items}(\mathcal{S})$
\WHILE{$\mathcal{P} \neq \emptyset$}
    \STATE $\mathcal{P}' \gets \emptyset$
    \STATE $r \gets 0$
    \FOR{each $(name, spec) \in \mathcal{P}$}
        \STATE $\mathcal{D} \gets \mathrm{Dependencies}(spec)$
        \IF{$\mathrm{AllResolved}(\mathcal{D}, \mathcal{R})$}
            \STATE $\mathcal{R}[name] \gets \mathrm{Resolve}(spec, \mathcal{R}, \mathcal{V}, s)$
            \STATE $r \gets r+1$
        \ELSE
            \STATE $\mathcal{P}' \gets \mathcal{P}' \cup \{(name, spec)\}$
        \ENDIF
    \ENDFOR
    \IF{$r=0$}
        \STATE \textbf{raise} unresolved dependency error
    \ENDIF
    \STATE $\mathcal{P} \gets \mathcal{P}'$
\ENDWHILE
\STATE $\mathcal{I} \gets \mathrm{ResolveInterfaces}(\mathcal{B}, \mathcal{R}, \mathcal{C})$
\STATE $\mathcal{X} \gets \mathrm{RenderArtifacts}(\mathcal{B}, \mathcal{R}, \mathcal{I})$
\STATE \textbf{return} executable task instance $\mathcal{X}$
\end{algorithmic}
\end{algorithm}

\subsection{Task Parameterization and Instantiation}
\label{sec:task_param}
Once a task has been validated, we transform it into a parameterized task template.
% This ordering is important: parameterization should be applied to a task whose behavior is already known to be correct, rather than to an unvalidated candidate task, where parameterization could amplify existing errors.
We abstract the varying fields of the validated task into a declarative parameter schema while preserving the original topology and task objective and keeping the test-case logic consistent with that objective.
The schema declares one parameter for each field intended to vary.
These fields include router names, link media, subnet allocations and prefix lengths, and protocol attributes such as administrative distances.
For each parameter, the schema specifies its domain of admissible values, its dependencies on other parameters, and any distinctness or non-overlap constraints it must satisfy.
Table~\ref{tab:template_example} in Appendix~\ref{app:template_example} summarizes the structure of a parameterized task template before rendering.

During task setup, the instantiation procedure resolves these parameters under explicit constraints rather than through unconstrained randomization.
The procedure also resolves physical interfaces from the emulator's capability specification, ensuring that the interface names in the generated configurations are compatible with the device model and the selected link-media parameters.
After all logical parameters and physical interfaces have been resolved, the procedure renders the template variables to generate mutually consistent topology specifications, initial configurations, reference configurations, and test-case definitions.
Algorithm~\ref{alg:task-instantiation} illustrates this process.

Parameterization is therefore intended to make direct reuse of memorized configuration snippets less effective; it does not by itself create fundamentally different operational scenarios.
Diversity across the benchmark instead originates from template-to-template variation, which spans different topologies, protocol features, and task objectives.

This pipeline preserves the realism and protocol diversity of existing configuration materials while converting them into executable benchmark tasks with consistent evaluation criteria.
The resulting benchmark contains 96 task templates covering representative IP routing, MPLS, and IP Overlay scenarios.
Table~\ref{tab:task_taxonomy} in Appendix~\ref{app:template_example} reports the full task taxonomy, including protocol categories, subcategories, representative features, template counts, and evaluation targets.

\section{Experiments and Findings}
\label{sec:evaluation}

% In this section, we describe the experimental setup and present the corresponding results.
Our experiments are organized around three research questions.
\textbf{RQ1:} How well do current LLM agents perform on network configuration tasks?
\textbf{RQ2:} How does agent behavior differ across models and thinking modes?
\textbf{RQ3:} What are the main causes of agent failures?

\begin{table*}[t]
  \centering
  \caption{Outcome and process metrics for evaluated LLM agents in \NetConfArena.}
  \label{tab:exp_main_results}
  \resizebox{\textwidth}{!}{
  \begin{tabular}{cccccccccccc}
    \toprule
    Model & Think & Score & Pass rate & Cfg. F1 & Turns & Err. act. & Rep. act. & Limit & Early sub. & Corr. sub. & Err. sub. \\
    \midrule
    \multirow{2}{*}{qwen3-8B} & \sNo & 0.337 & 0.046 & 0.508 & \underline{\textbf{18.26}} & \underline{\textbf{0.455}} & \underline{\textbf{0.496}} & \underline{\textbf{0.833}} & 0.000 & 0.046 & 0.121 \\
    & \sYes & 0.457 & 0.094 & 0.582 & 14.21 & 0.317 & 0.124 & 0.446 & 0.000 & 0.094 & \underline{\textbf{0.460}} \\
    \multirow{2}{*}{qwen3-32B} & \sNo & 0.656 & 0.233 & 0.684 & 15.61 & 0.090 & 0.017 & 0.473 & 0.000 & 0.154 & 0.373 \\
    & \sYes & 0.728 & 0.421 & 0.682 & 10.88 & 0.139 & 0.020 & 0.208 & 0.013 & 0.379 & 0.400 \\
    \multirow{2}{*}{deepseek-v4-flash} & \sNo & 0.914 & 0.656 & 0.759 & 14.72 & 0.027 & 0.000 & 0.350 & 0.000 & 0.502 & 0.148 \\
    & \sYes & 0.933 & 0.794 & 0.739 & 14.42 & 0.038 & 0.001 & 0.246 & 0.000 & 0.675 & 0.079 \\
    \multirow{2}{*}{deepseek-v4-pro} & \sNo & \underline{\textbf{0.961}} & \underline{\textbf{0.852}} & \underline{\textbf{0.796}} & 13.70 & 0.020 & 0.001 & 0.235 & 0.000 & \underline{\textbf{0.731}} & 0.033 \\
    & \sYes & 0.862 & 0.738 & 0.692 & 11.12 & 0.019 & 0.003 & 0.131 & \underline{\textbf{0.079}} & 0.692 & 0.098 \\
    \bottomrule
  \end{tabular}
  }
  \par\vspace{0.6em}
  \makebox[\textwidth][c]{\parbox{0.96\textwidth}{\footnotesize Note: \sNo denotes thinking disabled, and \sYes denotes thinking enabled.
  Score denotes test-case score, Pass rate denotes the task pass rate, Cfg. F1 denotes configuration F1, Limit denotes interaction-limit hit, and sub. denotes submission.
  Err. act. and Rep. act. denote error-action and repeat-action rates, respectively.
  Bold underlined values mark the maximum value in each column, although larger values are not always better.}}
\end{table*}

\subsection{Evaluation Setup}

We construct LLM agents and evaluate them under eight model--thinking-mode settings in \NetConfArena.
% The experimental settings are described below.

\subsubsection{LLMs}
We evaluate qwen3-\{8B, 32B\}~\cite{yang2025qwen3} and deepseek-v4-\{flash, pro\}~\cite{deepseek2026v4preview}.
These models cover both relatively small backbones that can be deployed locally and stronger models that are primarily accessed through APIs.
We intentionally include different model families in order to analyze how model-specific behavior affects task performance.
For each model, we evaluate two thinking modes: thinking disabled and thinking enabled~\cite{openai2024learning,aliyun2024qwqplus}.
% In the thinking-enabled mode, the LLM is allowed to generate longer intermediate reasoning before outputting the answer~\cite{openai2024learning,aliyun2024qwqplus}.
% In the thinking-disabled mode, the agent is instructed to produce the answer directly.
For all models and thinking modes, we set the decoding temperature to 0.

\subsubsection{Agent Framework}
All LLMs are evaluated under the same ReAct-style~\cite{yao2023react} workflow, which provides a consistent interaction interface and process structure.
As shown in Appendix~\ref{app:system_prompt}, the system prompt is fixed across models and thinking modes, while the task context follows the same template across task instances.
To evaluate whether agents can complete tasks within a bounded operation budget, we impose a maximum interaction limit.
Specifically, each run is limited to 20 interaction turns.

\subsubsection{Task Instances}
We generate five concrete task instances from each of the 96 validated task templates using fixed random seeds.
The same five instances of each template are used for all model and thinking-mode combinations.
Thus, each agent is evaluated on 480 task instances, and the full evaluation produces 3840 execution trajectories.

\subsection{Main Results and Analysis}
\label{sec:main_results}

\begin{figure}[t]
  \centering
  \includegraphics[width=\linewidth]{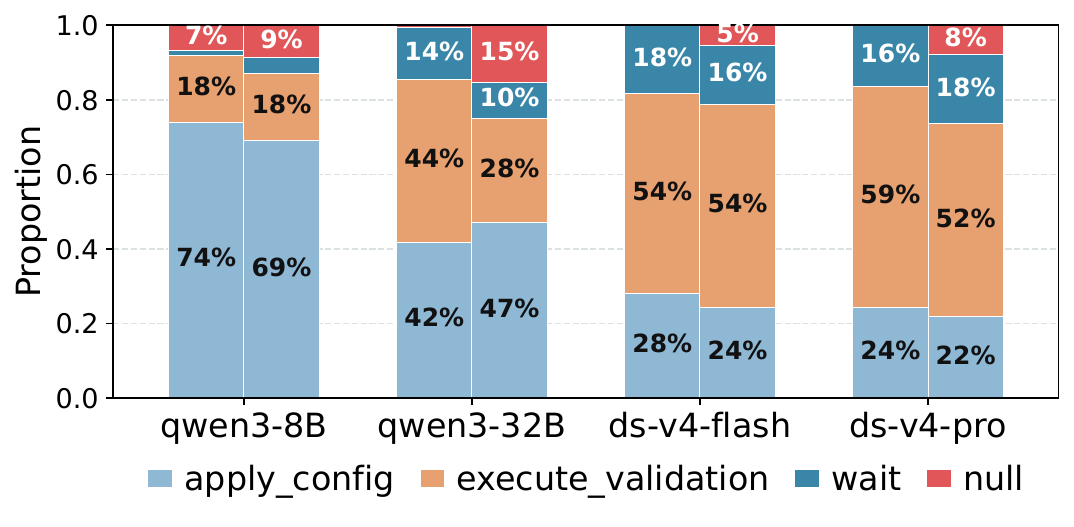}
  \caption{Action distribution across models and thinking modes.
  The distribution excludes \texttt{get\_running\_config} and \texttt{submit} actions.
  For each model, the left bar corresponds to thinking disabled and the right bar corresponds to thinking enabled.}
  \label{fig:exp_process_summary}
\end{figure}

Following the metrics defined in Section~\ref{sec:dual_perspective_eval}, Table~\ref{tab:exp_main_results} reports the mean values of the outcome metrics and main process metrics used for the main comparison.
Action distributions and task overhead are analyzed separately in Figures~\ref{fig:exp_process_summary} and~\ref{fig:exp_process_deep_dive}.

\subsubsection{Outcome-level analysis}

The test-case score measures whether the final network state satisfies task-specific test cases, while configuration F1 is reported as a complementary text-level reference metric.
Both metrics are computed per task instance and then averaged over the 480 instances with equal weight.
The task pass rate is the fraction of these instances with a full test-case score.

Table~\ref{tab:exp_main_results} shows a clear capability gap across model settings.
The strongest setting is deepseek-v4-pro with thinking disabled, which achieves a test-case score of 0.961, a task pass rate of 0.852, and a configuration F1 of 0.796.
The deepseek-v4-flash setting with thinking enabled also performs strongly, reaching a test-case score of 0.933 and a task pass rate of 0.794.
In contrast, qwen3-8B remains substantially weaker, with test-case scores of 0.337 and 0.457 and task pass rates of 0.046 and 0.094 under the thinking-disabled and thinking-enabled settings, respectively.
In every setting the task pass rate falls well below the test-case score, showing that partial credit masks many instances in which agents complete most but not all checkpoints.

Thinking improves the test-case score of qwen3-8B and qwen3-32B by 12.0 and 7.2 percentage points, respectively.
For deepseek-v4-flash, thinking brings only a small score improvement of 1.9 percentage points.
Interestingly, for deepseek-v4-pro, thinking reduces the test-case score by 9.9 percentage points.
As discussed in the process-level analysis below, this performance degradation is closely associated with an increase in early submission.

These results indicate that smaller LLMs still face substantial limitations in completing complex network configuration tasks.
Thinking can help weaker models by allowing them to reason through the task execution more explicitly.
However, stronger LLMs do not benefit uniformly from thinking; in some cases, thinking can even degrade final performance.
Appendix~\ref{app:generalizability} further examines how these scores vary across protocol categories and across instances of the same template.

\begin{center}
\fcolorbox{gray!45!black}{gray!8}{\begin{minipage}{0.96\linewidth}
\small
\emph{Answer to RQ1:} Agents built on state-of-the-art LLMs can complete most \NetConfArena tasks through interaction with the executable network environment, but their performance remains imperfect and setting-dependent.
Agents built on smaller models struggle substantially, even when thinking mode is enabled.
\end{minipage}}
\end{center}

\subsubsection{Process-level analysis}

% Process-level evaluation explains how agents progress through the closed-loop interaction before termination.

\paragraph{Action Distribution}
Figure~\ref{fig:exp_process_summary} shows the action distribution of the trajectories, revealing how agents allocate their interaction budget between different action types.
Different model settings exhibit distinct execution styles.
Stronger DeepSeek settings spend approximately 52--59\% of their actions on validation, whereas qwen3-8B spends a much larger proportion of its actions on configuration updates.
This suggests that stronger agents do not merely issue fewer commands; they also allocate more of their interaction budget to checking the executable network state.

Invalid or unrecognized action turns also consume part of the interaction budget without changing or inspecting the network state.
These turns are more visible in weaker or less stable settings, indicating that part of their lower efficiency comes from failing to produce executable tool calls rather than from configuration reasoning alone.
Across all evaluated models, enabling thinking increases the share of null or unrecognized action turns.
The increase is largest for qwen3-32B, where this share rises from 0.5\% to 13.0\%.
This suggests that thinking mode lengthens the interaction context and may make generated actions less likely to follow the required format.

\paragraph{Action Quality}
Table~\ref{tab:exp_main_results} shows that weaker settings have substantially higher error-action and repeat-action rates.
These metrics capture whether issued actions are executable and effective, rather than invalid or unrecognized tool calls.
They therefore more directly reflect the model's reasoning ability in a specific network environment and its command-level network configuration knowledge.

Because smaller models have limited capacity to encode and apply network configuration knowledge, they are more prone to generate ineffective configuration commands.
For example, qwen3-8B with thinking disabled exhibits frequent invalid or repeated actions.
By contrast, deepseek-v4-pro with thinking disabled has an error-action rate of only 2.0\% and almost no repeated actions.
For the qwen3-8B settings, enabling thinking reduces its error and repetition patterns, suggesting that additional reasoning helps smaller models produce more executable configuration actions.

\paragraph{Completion Mode}
Table~\ref{tab:exp_main_results} reports the completion-mode distribution.
Weaker settings are more likely to exhaust the interaction budget, indicating that they often fail to converge within the 20-turn limit.
When models actively submit, we further find that smaller models often fail to judge whether the task has truly been completed.
This appears as higher error-submission rates and lower correct-submission rates.
The DeepSeek settings are much stronger on this dimension, with error-submission rates of 3.3--14.8\%, compared with 12.1--46.0\% for qwen3-8B and 37.3--40.0\% for qwen3-32B.

In addition, we observe an important exception for deepseek-v4-pro: thinking increases early submission from 0\% to 7.9\%.
In other settings, this early submission behavior is rare, with only 1.3\% of qwen3-32B trajectories and no qwen3-8B or deepseek-v4-flash trajectories ending with early submission.
Manual inspection of the corresponding trajectories shows that the model sometimes generates multiple actions in a single turn, while the interface permits only one action per turn.
After receiving this error feedback, the model directly submits the task instead of repairing the action sequence.

\begin{figure}[t]
  \centering
  \includegraphics[width=\linewidth]{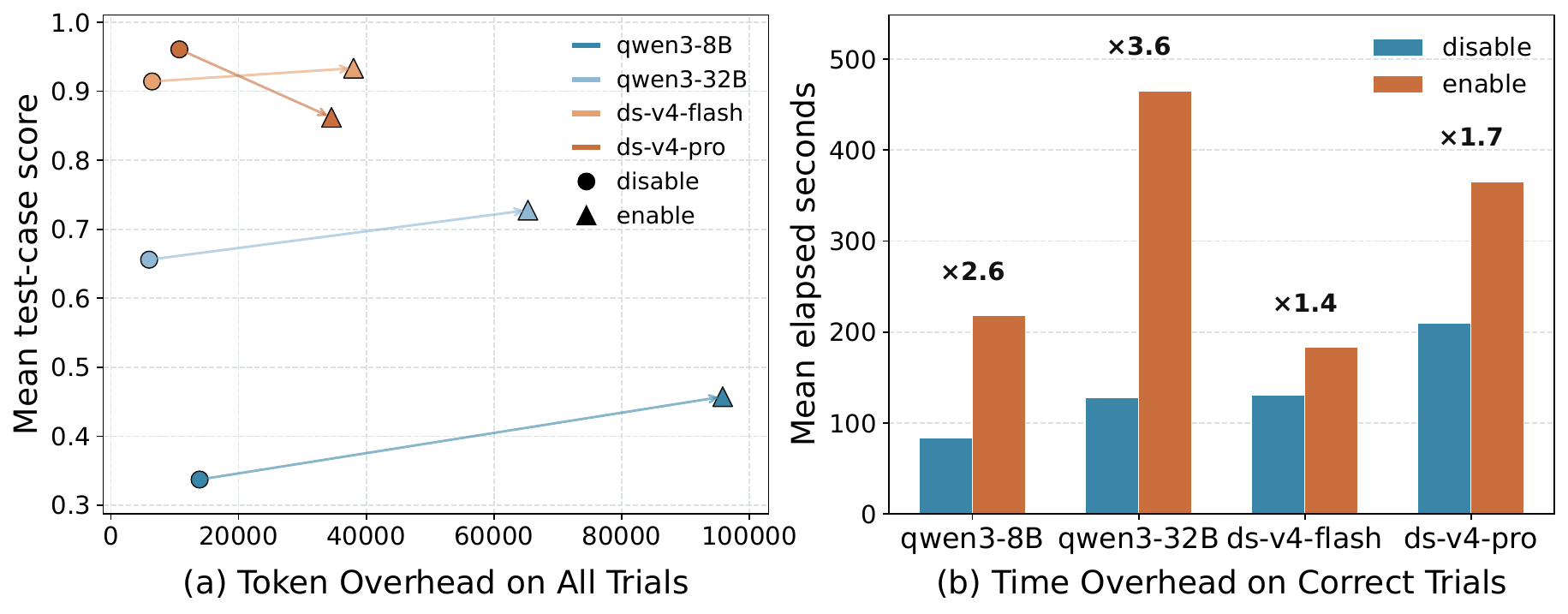}
  \caption{Task overhead comparison between thinking-disabled and thinking-enabled settings.}
  \label{fig:exp_process_deep_dive}
\end{figure}

\paragraph{Task Overhead}
For task-overhead metrics, we mainly examine whether the performance benefit of thinking mode is sufficient to justify its additional cost.
Figure~\ref{fig:exp_process_deep_dive} isolates the overhead introduced by thinking mode.
We use the number of assistant-output tokens to quantify reasoning overhead across model and thinking-mode settings.
Thinking substantially increases assistant-output token overhead for the two qwen3 models while improving test-case score.

We also compare time overhead between the thinking-disabled and thinking-enabled settings over trajectories that successfully complete the task, so that the comparison is not dominated by failed or prematurely terminated runs.
The time overhead of thinking mode is also especially pronounced for the qwen3 models.
By contrast, the DeepSeek models show a more moderate increase in time cost.

\begin{center}
\fcolorbox{gray!45!black}{gray!8}{\begin{minipage}{0.96\linewidth}
\small
\emph{Answer to RQ2:} Model choice and thinking mode affect process behavior.
Enabling thinking improves the qwen3 models and slightly improves deepseek-v4-flash, but it hurts deepseek-v4-pro by increasing early submission.
Specifically, enabling thinking trades fewer repeated or low-quality actions for higher interaction overhead and a greater risk of malformed actions.
\end{minipage}}
\end{center}

\subsection{Failure Taxonomy}
\label{sec:diagnosis}

Aggregate metrics characterize the behavior patterns of different agents, but they do not by themselves explain how failures occur.
We therefore use these metrics as diagnostic signals to guide further inspection of the corresponding trajectories, final configuration diffs, and evaluator feedback from failed test cases.
This analysis yields a taxonomy of four representative failure categories, all of which are made directly observable by the combination of realistic configuration tasks and process- and outcome-level evaluation in \NetConfArena.

\subsubsection{Knowledge deficiency}
This category captures cases in which the agent lacks the protocol, mechanism, or command knowledge needed to complete the task.
Such failures may appear as high-level conceptual mistakes, confusion between protocol-specific mechanisms, or misuse of device commands.
In RIP route summarization, for example, one failed trajectory applies a BGP- or OSPF-style workaround based on static blackhole routes and redistribution, rather than using RIP's interface-level \texttt{ip summary-address rip} command.
As a result, the expected summarization state is absent from test-case outputs, and additional redistribution changes also disturb the required RIP network statements.

\subsubsection{Unproductive hesitation}
This category captures cases in which the agent keeps observing or validating the network without converting feedback into effective configuration progress.
This pattern is reflected in runs that spend many turns on different validation actions, or repeatedly issue the same validation action, leaving too few turns to complete the remaining task subgoals.
In eBGP, for example, one failed trajectory repeatedly checks BGP summary output and advertised routes but never completes all required neighbor-establishment and route-advertisement subgoals.
The agent is interacting with the network, but its validation actions do not guide it toward a complete solution.

\subsubsection{Premature abandonment}
This category captures cases in which the agent voluntarily submits after a malformed action or unexpected observation, instead of adjusting its subsequent action.
These cases often reflect unstable reasoning when thinking is enabled: the agent may emit several actions in one turn and infer observations it never received.
Instead of using the error signal to recover, it abandons the task and submits.
In BGP Peer Group, for example, one run terminates after only two turns: the first turn is parsed as null, and the second turn submits with no configuration attempt.
Other model settings solve the same task with full test-case scores, suggesting that the failure is not due to task difficulty but premature loss of interaction control.

\subsubsection{Specification deviation}
This category captures cases in which the agent rewrites task-provided invariants, such as addresses, prefixes, AS numbers, area identifiers, or required mechanisms.
The resulting configuration may be internally coherent, but it no longer solves the specified task instance.
Such failures suggest that the agent may rely on familiar configuration patterns learned from prior data rather than grounding its actions in the task-specific parameters.
In LDP VPLS, for example, one trajectory replaces the given 192.168.x link addresses with a self-designed 10.x /30 addressing scheme.
Although the configuration remains locally reasonable, test cases tied to the specified next hops fail because the task parameters have been changed.

\begin{center}
\fcolorbox{gray!45!black}{gray!8}{\begin{minipage}{0.96\linewidth}
\small
\emph{Answer to RQ3:} The failure taxonomy shows that failures are not caused only by knowledge deficiencies.
Specification deviation, unproductive hesitation, and premature abandonment further indicate that current agents remain unreliable in planning and executing network configuration tasks.
\end{minipage}}
\end{center}

\section{Discussion}
\label{sec:future_directions}

\subsection{Directions for Enhancing LLM Agents}
\label{sec:enhancing_agents}

\NetConfArena can support future improvements in two complementary ways: trajectory-based learning to improve foundation models and harness-level mechanisms to manage the interaction around the model.

\subsubsection{Learning from Validated Trajectories}

\NetConfArena records complete interaction traces that connect the task context with the agent's execution process and final test-case score.
These traces can be converted into supervision for fine-tuning LLM agents~\cite{yang2024swe}.

The most direct use is to select high-quality successful trajectories as demonstrations.
Because \NetConfArena evaluates final behavior rather than textual similarity to a single reference configuration, this selection can preserve different valid strategies for the same operational intent.
Moreover, partial and failed trajectories can support preference-based training, using test-case scores and process metrics as automatically derived preference signals~\cite{rafailov2023dpo}.

\subsubsection{Improving the Agent Harness}

% A second direction is to strengthen the harness that mediates between the LLM agent and the executable network.
The failures analyzed in Section~\ref{sec:diagnosis} indicate that many errors cannot be attributed solely to gaps in the model's protocol knowledge; they also stem from weaknesses in how the interaction process itself is managed across multiple turns.

Skills offer a practical means of supplying the procedural knowledge that network configuration tasks demand.
A skill can encode a reusable procedure for a given protocol feature, bundling prerequisite checks, command templates, validation commands, and common repair rules.
Beyond this, the harness can maintain an explicit summary of the task state and use it to keep the agent aligned with unresolved subgoals.
It can also enforce safety constraints prior to execution by checking whether a proposed action overwrites task-provided addresses, prefixes, AS numbers, or other invariants.

\subsection{Limitations and Future Work}
\label{sec:limitations}

\NetConfArena has two main limitations.

\begin{itemize}
    \item \textbf{Task realism.}
    The tasks in \NetConfArena are controlled configuration scenarios on small topologies, a design that keeps evaluation reproducible but omits much of the complexity of production networks.
    Concerns such as rollback procedures and multi-vendor environments therefore fall outside the current scope of the benchmark.
    \item \textbf{Evaluation scope.}
    Our evaluation focuses on LLM agents, so traditional configuration synthesis tools are not included as baselines; they start from formal specifications rather than from natural-language intent.
    Our metrics likewise target the functional correctness of the final network behavior rather than network performance such as latency, congestion, or transient routing dynamics.
\end{itemize}

Future work will extend the benchmark along both the task and evaluation dimensions.
On the task dimension, we plan to add tasks that begin from legacy configurations conflicting with the intended objective, as well as tasks that target fault-tolerant configuration scenarios.
On the evaluation dimension, we plan to broaden the coverage of our test cases through integration with formal network verification tools.

\section{Related Work}

\subsection{LLMs for Network Configuration}

Recent work uses LLMs for network configuration generation, intent translation, and operator assistance~\cite{mondal2023llms,liu2025cegs,lian2024large,jiang2024caip,ifland2024genet,jeong2024s,mani2023enhancing}.
\NetConfArena differs by evaluating interactive agents in executable environments, where agents must use network feedback rather than only producing text.
This focus is complementary to network verification and synthesis.
Verification tools check whether configurations satisfy formal network properties~\cite{fogel2015general,beckett2017general,gember2016fast,brown2023lessons}, while synthesis systems generate configurations from higher-level specifications~\cite{beckett2016don,beckett2017network,el2018netcomplete,ramanathan2023practical}.
\NetConfArena can incorporate such tools as test cases or feedback sources, but its goal is to evaluate whether LLM agents can accomplish configuration through closed-loop interaction.

\subsection{Benchmarks for LLM Agents}

LLM agent benchmarks span domains such as web navigation, software engineering, tool use, and office automation~\cite{yao2023react,yang2024swe,lai2024autowebglm,zhang2025webpilot,hong2023metagpt,liu2025pc}.
Existing LLM benchmarks for network configuration largely focus on intent-to-command translation or simplified configuration tasks with limited interaction and reduced environmental fidelity~\cite{wang2024netconfeval,aykurt2024netllmbench,zhou2026netarena}.
This simplification leaves important domain requirements underexplored: realistic network configuration is stateful and protocol-dependent, and command acceptance does not guarantee correct network behavior.
These properties call for benchmarks that expose agents to executable feedback and evaluate whether their actions produce the intended network behavior.
\NetConfArena addresses this gap by treating network configuration as an interactive decision-making problem and measuring both final behavior and intermediate process quality.

\section{Conclusion}

We presented \NetConfArena, an executable benchmark for evaluating LLM agents in closed-loop network configuration.
\NetConfArena combines emulator-backed interaction, executable test cases, and process-level diagnosis to assess LLM agents in terms of task success and execution reliability.
We have released the benchmark framework and task suite as open source.
% offering a grounded way to study LLM agents on multi-device, protocol-rich configuration tasks

\bibliographystyle{IEEEtran}
\bibliography{ref}

\clearpage
\appendix
\subsection{Agent Prompt for Evaluation}
\label{app:system_prompt}

All evaluated models use the same network-operator prompt wrapper.
The wrapper has two parts: a fixed system prompt that defines the agent role and tool protocol, and a parameterized task prompt that injects the concrete scenario, topology, and goals for one benchmark instance.
The system prompt states that tool calls return JSON-formatted observations, exposes actions for configuration, validation, convergence waiting, and final submission, and restricts each turn to exactly one executable \texttt{Action} with an optional \texttt{Thought}.
The following box shows the prompt in a compact form.

\begin{center}
\setlength{\fboxsep}{4pt}
\fbox{%
\begin{minipage}{0.96\columnwidth}
\scriptsize
{\ttfamily
\textbf{System prompt.}\\
You are an intelligent agent for network operation, skilled at configuring networks based on the requirements I provide.\\
I will give you different task scenarios, network topologies, and the goals you need to achieve.\\
Complete the goals in a logical order, ensuring each goal is fully achieved before moving to the next.\\
Each tool returns a JSON-formatted response.\\
Tools: \\
\# get\_running\_config(<devices\_names>)\\
\# apply\_config(<config\_commands>)\\
\# execute\_validation(<validation\_command>)\\
\# wait(<seconds>)\\
\# submit()\\
The device-name argument is a list of unique hostnames.\\
Configuration and validation arguments are dictionaries from device names to newline-separated command strings.\\
Validation commands must be read-only and should not change device state.\\
Each response may include a Thought line and must include exactly one executable Action line.\\
Reply with only one action at a time.\\[0.25em]
Few-shot examples: get\_running\_config(["Router"]); apply\_config(\{"router-a": "interface Gi0/0\textbackslash{}nno shutdown"\}); \ldots; submit().\\[0.25em]
\textbf{Task prompt.}\\
The current task you need to perform is as follows:\\
\# Scenario\\
\$\{scenario\}\\
\# Topology\\
\$\{topology\}\\
\# Goals\\
\$\{goals\}
}
\end{minipage}}
\end{center}

\subsection{Full Trajectory Example}
\label{app:trajectory_example}

This appendix expands the successful execution of Fig.~\ref{fig:motivating}(b) into its full 19-turn trajectory, which passes all hidden test cases of the task instance.
The topology and the three per-prefix objectives are those of Fig.~\ref{fig:motivating}(a).
The Core-facing links connect Aria and Brick to Core over \texttt{192.168.12.0/24} and \texttt{192.168.34.0/24}, respectively.
The lower links connect Aria and Brick to Eason over \texttt{192.168.13.0/24} and \texttt{192.168.24.0/24}, respectively.
Each turn below gives the agent's \texttt{Thought} and \texttt{Action} together with the \texttt{Observation} returned by the environment, abbreviated where device output is repetitive.

\begin{framed}
\scriptsize
\raggedright
\textbf{Turn 1.}\\*
\textit{Thought.} I first need the interface addresses and existing RIP configuration on all four routers.\\*
\textit{Action.} \texttt{get\_running\_config(["Aria", "Core", "Eason", "Brick"])}\\*
\textit{Observation.} \texttt{\{"Core": "... Loopback0 1.1.1.1/24\nn Loopback1 11.11.11.11/24\nn Loopback2 111.111.111.111/24\nn Ethernet0/0 192.168.12.2/24\nn Ethernet0/1 192.168.34.2/24\nn router rip\nn network 192.168.12.0\nn network 192.168.34.0", "Aria": "... Ethernet0/1 192.168.12.1/24\nn Ethernet0/0 192.168.13.1/24 ...", "Eason": "... Ethernet0/1 192.168.13.2/24\nn Ethernet0/0 192.168.24.2/24 ...", "Brick": "... Ethernet0/0 192.168.34.1/24\nn Ethernet0/1 192.168.24.1/24 ..."\}}\\[0.35em]
\textbf{Turn 2.}\\*
\textit{Thought.} The routers already advertise the inter-router links, but Core's three loopbacks are absent from RIP.\\*
I will enable RIPv2 without auto-summary everywhere and advertise the loopbacks from Core.\\*
\textit{Action.} \texttt{apply\_config(\{"Aria": "router rip\nn version 2\nn no auto-summary", "Core": "router rip\nn version 2\nn no auto-summary\nn network 1.0.0.0\nn network 11.0.0.0\nn network 111.0.0.0", "Eason": "router rip\nn version 2\nn no auto-summary", "Brick": "router rip\nn version 2\nn no auto-summary"\})}\\*
\textit{Observation.} \texttt{\{"Core": "Core(config-router)\#version 2\nn Core(config-router)\#no auto-summary\nn Core(config-router)\#network 1.0.0.0\nn Core(config-router)\#network 11.0.0.0\nn Core(config-router)\#network 111.0.0.0\nn Core(config-router)\#end", "Aria": "... Aria(config-router)\#end", "Eason": "...", "Brick": "..."\}}\\[0.35em]
\textbf{Turn 3.}\\*
\textit{Thought.} To make every path to 1.1.1.0/24 enter Core through Ethernet0/1, I will penalize the update that Core sends toward Aria through Ethernet0/0.\\*
\textit{Action.} \texttt{apply\_config(\{"Core": "ip access-list standard P1\nn permit 1.1.1.0 0.0.0.255\nn offset-list P1 out 5 Ethernet0/0"\})}\\*
\textit{Observation.} \texttt{\{"Core": "Core(config)\#ip access-list standard P1\nn Core(config-std-nacl)\#permit 1.1.1.0 0.0.0.255\nn Core(config-std-nacl)\#offset-list P1 out 5 Ethernet0/0\nn \% Invalid input detected at '\textasciicircum' marker.\nn Core(config-std-nacl)\#end"\}}\\[0.35em]
\textbf{Turn 4.}\\*
\textit{Thought.} The ACL was accepted, but the offset-list was rejected because I issued it from ACL configuration mode.\\*
I will bind P1 under \texttt{router rip} and also penalize the direct Core update that Brick receives for 11.11.11.0/24.\\*
\textit{Action.} \texttt{apply\_config(\{"Core": "router rip\nn offset-list P1 out 5 Ethernet0/0", "Brick": "ip access-list standard P2\nn permit 11.11.11.0 0.0.0.255\nn router rip\nn offset-list P2 in 5 Ethernet0/0"\})}\\*
\textit{Observation.} \texttt{\{"Core": "Core(config-router)\#offset-list P1 out 5 Ethernet0/0\nn Core(config-router)\#end", "Brick": "Brick(config)\#ip access-list standard P2\nn Brick(config-std-nacl)\#permit 11.11.11.0 0.0.0.255\nn Brick(config)\#router rip\nn Brick(config-router)\#offset-list P2 in 5 Ethernet0/0\nn Brick(config-router)\#end"\}}\\[0.35em]
\textbf{Turn 5.}\\*
\textit{Thought.} I initially try to make 111.111.111.0/24 unreachable from Eason by suppressing the update that Core sends toward Aria.\\*
\textit{Action.} \texttt{apply\_config(\{"Core": "ip access-list standard P3\nn permit 111.111.111.0 0.0.0.255\nn router rip\nn offset-list P3 out 16 Ethernet0/0"\})}\\*
\textit{Observation.} \texttt{\{"Core": "Core(config)\#ip access-list standard P3\nn Core(config-std-nacl)\#permit 111.111.111.0 0.0.0.255\nn Core(config)\#router rip\nn Core(config-router)\#offset-list P3 out 16 Ethernet0/0\nn Core(config-router)\#end"\}}\\[0.35em]
\textbf{Turn 6.}\\*
\textit{Thought.} RIP needs time to propagate the changed metrics before I evaluate the policies.\\*
\textit{Action.} \texttt{wait(30)}\\*
\textit{Observation.} \texttt{\{"status": "30 seconds elapsed"\}}\\[0.35em]
\textbf{Turn 7.}\\*
\textit{Thought.} I will first check the two path-selection objectives.\\*
\textit{Action.} \texttt{execute\_validation(\{"Aria": "show ip route 1.1.1.0", "Brick": "show ip route 11.11.11.0"\})}\\*
\textit{Observation.} \texttt{\{"Aria": "R 1.1.1.0 [120/1] via 192.168.12.2, Ethernet0/1", "Brick": "R 11.11.11.0 [120/3] via 192.168.24.2, Ethernet0/1"\}}\\[0.35em]
\textbf{Turn 8.}\\*
\textit{Thought.} P2 works, but P1 is missing because Aria still uses its direct Core link.\\*
I will also check whether P3 produced the required selective reachability.\\*
\textit{Action.} \texttt{execute\_validation(\{"Aria": "show ip route 111.111.111.0", "Eason": "show ip route 111.111.111.0", "Brick": "show ip route 111.111.111.0"\})}\\*
\textit{Observation.} \texttt{\{"Aria": "\% Subnet not in table", "Eason": "R 111.111.111.0 [120/2] via 192.168.24.1, Ethernet0/0", "Brick": "R 111.111.111.0 [120/1] via 192.168.34.2, Ethernet0/0"\}}\\[0.35em]
\textbf{Turn 9.}\\*
\textit{Thought.} P3 blocks the wrong router, while Eason still learns the prefix through Brick.\\*
I will inspect Core to determine why P1 disappeared.\\*
\textit{Action.} \texttt{get\_running\_config(["Core"])}\\*
\textit{Observation.} \texttt{\{"Core": "... ip access-list standard P1\nn permit 1.1.1.0 0.0.0.255\nn ip access-list standard P3\nn permit 111.111.111.0 0.0.0.255\nn router rip\nn offset-list P3 out 16 Ethernet0/0\nn network 1.0.0.0\nn network 11.0.0.0\nn network 111.0.0.0 ..."\}}\\[0.35em]
\textbf{Turn 10.}\\*
\textit{Thought.} P3 replaced P1 because both offset-lists use the same interface and direction on Core.\\*
I will remove P3 from Core and restore P1 before moving the reachability policy to Eason.\\*
\textit{Action.} \texttt{apply\_config(\{"Core": "router rip\nn no offset-list P3 out 16 Ethernet0/0\nn offset-list P1 out 5 Ethernet0/0"\})}\\*
\textit{Observation.} \texttt{\{"Core": "Core(config-router)\#no offset-list P3 out 16 Ethernet0/0\nn Core(config-router)\#offset-list P1 out 5 Ethernet0/0\nn Core(config-router)\#end"\}}\\[0.35em]
\textbf{Turn 11.}\\*
\textit{Thought.} I will verify that the restored P1 sends every remote router toward Core Ethernet0/1.\\*
\textit{Action.} \texttt{execute\_validation(\{"Aria": "show ip route 1.1.1.0", "Eason": "show ip route 1.1.1.0", "Brick": "show ip route 1.1.1.0"\})}\\*
\textit{Observation.} \texttt{\{"Aria": "R 1.1.1.0 [120/3] via 192.168.13.2, Ethernet0/0", "Eason": "R 1.1.1.0 [120/2] via 192.168.24.1, Ethernet0/0", "Brick": "R 1.1.1.0 [120/1] via 192.168.34.2, Ethernet0/0"\}}\\[0.35em]
\textbf{Turn 12.}\\*
\textit{Thought.} Eason can learn P3 through either neighbor, so I will filter the update received from Aria first.\\*
\textit{Action.} \texttt{apply\_config(\{"Eason": "ip access-list standard P3\nn permit 111.111.111.0 0.0.0.255\nn interface Ethernet0/1\nn ip rip offset-list P3 in 16"\})}\\*
\textit{Observation.} \texttt{\{"Eason": "Eason(config)\#ip access-list standard P3\nn Eason(config-std-nacl)\#permit 111.111.111.0 0.0.0.255\nn Eason(config)\#interface Ethernet0/1\nn Eason(config-if)\#ip rip offset-list P3 in 16\nn \% Invalid input detected at '\textasciicircum' marker.\nn Eason(config-if)\#end"\}}\\[0.35em]
\textbf{Turn 13.}\\*
\textit{Thought.} The interface-level form is unsupported, so the offset must be configured under \texttt{router rip} with the interface as an argument.\\*
\textit{Action.} \texttt{apply\_config(\{"Eason": "router rip\nn offset-list P3 in 16 Ethernet0/1"\})}\\*
\textit{Observation.} \texttt{\{"Eason": "Eason(config-router)\#offset-list P3 in 16 Ethernet0/1\nn Eason(config-router)\#end"\}}\\[0.35em]
\textbf{Turn 14.}\\*
\textit{Thought.} I must also cover the update Eason receives from Brick through Ethernet0/0.\\*
\textit{Action.} \texttt{apply\_config(\{"Eason": "router rip\nn offset-list P3 in 16 Ethernet0/0"\})}\\*
\textit{Observation.} \texttt{\{"Eason": "Eason(config-router)\#offset-list P3 in 16 Ethernet0/0\nn Eason(config-router)\#end"\}}\\[0.35em]
\textbf{Turn 15.}\\*
\textit{Thought.} I will allow both incoming metric changes to converge before checking reachability.\\*
\textit{Action.} \texttt{wait(30)}\\*
\textit{Observation.} \texttt{\{"status": "30 seconds elapsed"\}}\\[0.35em]
\textbf{Turn 16.}\\*
\textit{Thought.} I will verify P3 and audit the final offset-list placement for another collision.\\*
\textit{Action.} \texttt{execute\_validation(\{"Aria": "show ip route 111.111.111.0", "Brick": "show ip route 111.111.111.0", "Eason": "show ip route 111.111.111.0\nn show running-config | include offset-list"\})}\\*
\textit{Observation.} \texttt{\{"Aria": "R 111.111.111.0 [120/1] via 192.168.12.2, Ethernet0/1", "Brick": "R 111.111.111.0 [120/1] via 192.168.34.2, Ethernet0/0", "Eason": "\% Subnet not in table\nn offset-list P3 in 16 Ethernet0/1\nn offset-list P3 in 16 Ethernet0/0"\}}\\[0.35em]
\textbf{Turn 17.}\\*
\textit{Thought.} P3 now holds, and its two bindings occupy distinct interfaces.\\*
I will revalidate the complete forwarding paths for P1 and P2.\\*
\textit{Action.} \texttt{execute\_validation(\{"Aria": "show ip route 1.1.1.0", "Eason": "show ip route 1.1.1.0", "Brick": "show ip route 1.1.1.0\nn show ip route 11.11.11.0"\})}\\*
\textit{Observation.} \texttt{\{"Aria": "R 1.1.1.0 [120/3] via 192.168.13.2, Ethernet0/0", "Eason": "R 1.1.1.0 [120/2] via 192.168.24.1, Ethernet0/0", "Brick": "R 1.1.1.0 [120/1] via 192.168.34.2, Ethernet0/0\nn R 11.11.11.0 [120/3] via 192.168.24.2, Ethernet0/1"\}}\\[0.35em]
\textbf{Turn 18.}\\*
\textit{Thought.} The route tables satisfy both path objectives, so I will confirm the selective reachability objective with data-plane probes.\\*
\textit{Action.} \texttt{execute\_validation(\{"Aria": "ping 111.111.111.111", "Brick": "ping 111.111.111.111", "Eason": "ping 111.111.111.111"\})}\\*
\textit{Observation.} \texttt{\{"Aria": "!!!!!\nn Success rate is 100 percent (5/5)", "Brick": "!!!!!\nn Success rate is 100 percent (5/5)", "Eason": ".....\nn Success rate is 0 percent (0/5)"\}}\\[0.35em]
\textbf{Turn 19.}\\*
\textit{Thought.} All three objectives are satisfied after repairing the rejected commands and the offset-list collision.\\*
\textit{Action.} \texttt{submit()}
\end{framed}

\subsection{Benchmark Details}
\label{app:template_example}
\label{app:metric_details}

This appendix provides additional details on the benchmark tasks, task templates, and evaluation test cases used in \NetConfArena.

Each \NetConfArena task template contains a natural-language intent, a topology specification, initial device configurations, a reference configuration, and executable test cases.
The reference configuration is used for configuration-level analysis, but it is not the sole correctness criterion because multiple command sequences may produce the same intended behavior.
Table~\ref{tab:template_example} summarizes the structure of a parameterized task template before rendering, including its intent, topology, configurations, test cases, and parameter constraints.
The example is based on a BGP Local Preference and MED template covering iBGP and eBGP session establishment, route advertisement and reachability, and inbound and outbound path preference.
The complete template contains the fields shown in the table together with all router-specific interface bindings and generated test cases.

Test cases are defined around sub-intents, such as route advertisement, selected next hop, neighbor state, or protocol counters.
This allows the benchmark to report partial completion and to diagnose which aspect of the operator intent was violated.
Table~\ref{tab:testcase_example} illustrates this design by mapping representative intents from the BGP path-control template to device-level diagnostic commands and deterministic predicates.

\begin{table}[!t]
  \centering
  \caption{Structure of a parameterized task template before rendering.}
  \label{tab:template_example}
  \scriptsize
  \setlength{\tabcolsep}{3pt}
  \renewcommand{\arraystretch}{1.05}
  \begin{tabular}{p{0.28\columnwidth}p{0.63\columnwidth}}
    \toprule
    Field & Template-level content \\
    \midrule
    Scenario &
    Defines the task category, e.g., BGP Local Preference and MED. \\
    Goals &
    Specifies the natural-language intent with placeholders, including iBGP over loopbacks, eBGP sessions, loopback advertisement, reachability, MED-based inbound preference, and Local Preference-based outbound preference. \\
    Topology model &
    Describes abstract nodes and links before rendering, including four routers and five logical links whose media are parameterized. \\
    Startup configurations &
    Defines initial per-router configuration templates using placeholders for interfaces, loopbacks, link addresses, masks, and EIGRP networks. \\
    Reference configurations &
    Defines reference configuration templates for BGP neighbors, update sources, advertised loopbacks, route maps, MED, and Local Preference. \\
    Test cases &
    Defines diagnostic commands and expected structured outcomes for BGP sessions, route advertisements, path selection, and policy effects. \\
    Parameters &
    Declares symbolic variables for router names, link media, subnets, loopback prefixes, addresses, masks, wildcards, and formatted CIDR strings. \\
    Sampling rules &
    Uses typed generators such as \texttt{choice}, \texttt{random\_subnet}, \texttt{subnet\_attr}, and \texttt{format}. \\
    Constraints &
    Enforces router-name uniqueness through \texttt{different\_from} constraints and prevents address conflicts through \texttt{no\_overlap} subnet constraints. \\
    \bottomrule
  \end{tabular}
\end{table}

\begin{table*}
  \centering
  \caption{Representative executable test cases for the BGP path-control task template.}
  \label{tab:testcase_example}
  \scriptsize
  \setlength{\tabcolsep}{4pt}
  \renewcommand{\arraystretch}{1.05}
  \begin{tabularx}{\textwidth}{
    >{\raggedright\arraybackslash}p{0.23\textwidth}
    >{\raggedright\arraybackslash}p{0.14\textwidth}
    >{\raggedright\arraybackslash}p{0.20\textwidth}
    >{\raggedright\arraybackslash}X
  }
    \toprule
    Intent & Device & Diagnostic command & Deterministic predicate \\
    \midrule

    iBGP sessions use loopback endpoints inside AS100
    & \texttt{router4\_name}
    & \texttt{show ip bgp neighbors}
    & Output contains established AS100 neighbors for \texttt{router2\_loopback0\_ip} and \texttt{router1\_loopback0\_ip}, with the local endpoint equal to \texttt{router4\_loopback0\_ip}. \\

    \midrule

    eBGP session between \texttt{router2\_name} and \texttt{router3\_name}
    & \texttt{router2\_name}
    & \texttt{show ip bgp neighbors}
    & Neighbor \texttt{link5\_ip2} is established with remote AS \texttt{200}, local IP \texttt{link5\_ip1}, and remote IP \texttt{link5\_ip2}. \\

    \midrule

    eBGP session between \texttt{router1\_name} and \texttt{router3\_name}
    & \texttt{router1\_name}
    & \texttt{show ip bgp neighbors}
    & Neighbor \texttt{link3\_ip2} is established with remote AS \texttt{200}, local IP \texttt{link3\_ip1}, and remote IP \texttt{link3\_ip2}. \\

    \midrule

    Full reachability to all loopback networks
    & \texttt{router3\_name}
    & \texttt{show ip bgp}
    & BGP table contains \texttt{router1\_loopback0\_network/24}, \texttt{router2\_loopback0\_network/24}, \texttt{router3\_loopback0\_network/24}, and \texttt{router4\_loopback0\_network/24}. \\

    \midrule

    MED policy is configured on AS100 eBGP exits
    & \texttt{router1\_name}, \texttt{router2\_name}
    & \texttt{show route-map}
    & Outbound route maps attached to \texttt{link3\_ip2} and \texttt{link5\_ip2} contain \texttt{set metric} clauses. \\

    \midrule

    AS100-bound traffic prefers the \texttt{router1\_name}--\texttt{router3\_name} link
    & \texttt{router3\_name}
    & \texttt{show ip bgp}
    & Best paths to AS100 loopback prefixes use next hop \texttt{link3\_ip1}, reflecting the lower MED advertised on the \texttt{router1\_name}--\texttt{router3\_name} eBGP session. \\

    \midrule

    Outbound traffic from AS100 prefers the \texttt{router1\_name}--\texttt{router3\_name} link
    & \texttt{router2\_name}, \texttt{router4\_name}
    & \texttt{show ip bgp}
    & Best path to \texttt{router3\_loopback0\_network/24} uses next hop \texttt{router1\_loopback0\_ip}, reflecting the higher Local Preference learned through \texttt{router1\_name}. \\

    \bottomrule
  \end{tabularx}
\end{table*}

The task suite covers protocol-related configuration scenarios across IP routing, MPLS, and IP Overlay\@.
These categories exercise different forms of network configuration reasoning, including path construction, neighbor establishment, route propagation, policy filtering, label distribution, and overlay control.
Table~\ref{tab:task_taxonomy} summarizes the 96 benchmark templates across 6 categories and 16 subcategories.
For each subcategory, the table reports the associated features, template count, and behavior-level evaluation targets.
Additional benchmark statistics are available in the accompanying repository.

\begin{table*}
  \centering
  \caption{Task taxonomy in \NetConfArena.}
  \label{tab:task_taxonomy}
  \scriptsize
  \setlength{\tabcolsep}{3pt}
  \renewcommand{\arraystretch}{1.08}
  \begin{tabularx}{\textwidth}{
    >{\raggedright\arraybackslash}p{1.35cm}
    >{\raggedright\arraybackslash}p{2.15cm}
    >{\raggedright\arraybackslash}X
    >{\centering\arraybackslash}p{1.8cm}
    >{\raggedright\arraybackslash}X
  }
    \toprule
    Category & Subcategory & Feature & \# Templates & Evaluation Target \\
    \midrule
    Static Routing & Basic Routing & Static, default, and floating routes & 1 & Route entries, next hops, AD-based failover, reachability \\
    \midrule
    \multirow{2}{*}{RIP} & Fundamentals & RIPv2 unicast neighbors & 1 & Neighbor peering, passive interfaces, no multicast/broadcast updates \\
    & Policy and Route Control & Summarization, offset-list, distribute-list, prefix-list, ACL & 3 & Aggregation, filtering, metric manipulation, path preference \\
    \midrule
    \multirow{3}{*}{OSPF} & Fundamentals & Single-area operation, network statements, router ID, interface cost & 1 & Adjacency, route exchange, network-statement forms, cost \\
    & Areas and Route Control & Stub/NSSA areas, summarization, DR/BDR, reference bandwidth & 8 & Area behavior, LSA visibility, summary routes, election and metric outcomes \\
    & Advanced Area Design & Virtual links, forward-address suppression, Type-7/5 translation & 4 & Backbone continuity, transit-area behavior, translator selection \\
    \midrule
    \multirow{4}{*}{BGP} & Peering, Propagation, and Scaling & iBGP/eBGP, update-source, transit, synchronization, soft reconfiguration, peer groups, route reflection, confederations & 12 & Session state, route propagation, next-hop reachability, scaling behavior \\
    & Routing Policy and Filtering & Private-AS removal, AS limits, prefix/ACL/AS-path filters, communities, allow-AS-in & 11 & Route acceptance, export policy, community handling, loop-prevention exceptions \\
    & Best-Path Selection and Traffic Engineering & Local AS, local preference, MED, weight, origin, AS path, router ID, DMZ bandwidth, backdoor & 13 & Best-path decisions, attribute manipulation, path preference, traffic steering \\
    & Advertisement Control and Aggregation & Unsuppress-map, conditional advertisement, aggregation, AS\_SET & 5 & Conditional export, suppression, summary generation, path-information retention \\
    \midrule
    \multirow{4}{*}{MPLS} & LDP and Label Forwarding & LDP sessions, label policy, explicit null, static binding, TTL and label ranges & 12 & Adjacency, bindings, LFIB forwarding, session protection, label behavior \\
    & Traffic Engineering & RSVP-TE, explicit paths, bandwidth admission, affinity, autoroute, FRR & 6 & CR-LSP setup, path constraints, reservations, protected forwarding \\
    & L3VPN & VRF/RD/RT, VPNv4, PE-CE routing, hub-and-spoke, route reflection, Inter-AS, GRE-core transport & 9 & Route import/export, VPN isolation, control-plane and data-plane reachability \\
    & L2VPN & VPWS, EoMPLS, AToM, pseudowire classes, Ethernet interworking, service aggregation & 4 & AC/PW binding, VC state, point-to-point L2 reachability, service isolation \\
    \midrule
    \multirow{2}{*}{IP Overlay} & IP Tunnels & L2TPv3, GRE, keepalive, MTU/MSS adjustment, IPsec tunnel protection & 4 & Tunnel and pseudowire state, overlay routing, fragmentation control, encryption \\
    & Dynamic Multipoint Overlay & DMVPN phases 1/2, mGRE, NHRP & 2 & Registration, hub relaying, shortcut forwarding, next-hop preservation \\
    \bottomrule
  \end{tabularx}
\end{table*}

\subsection{Generalizability Analysis}
\label{app:generalizability}

This section adds a new research question:
\textbf{RQ4:} Does agent performance generalize across task categories and different instances of the same task template?

\begin{figure}[t]
  \centering
  \includegraphics[width=\linewidth]{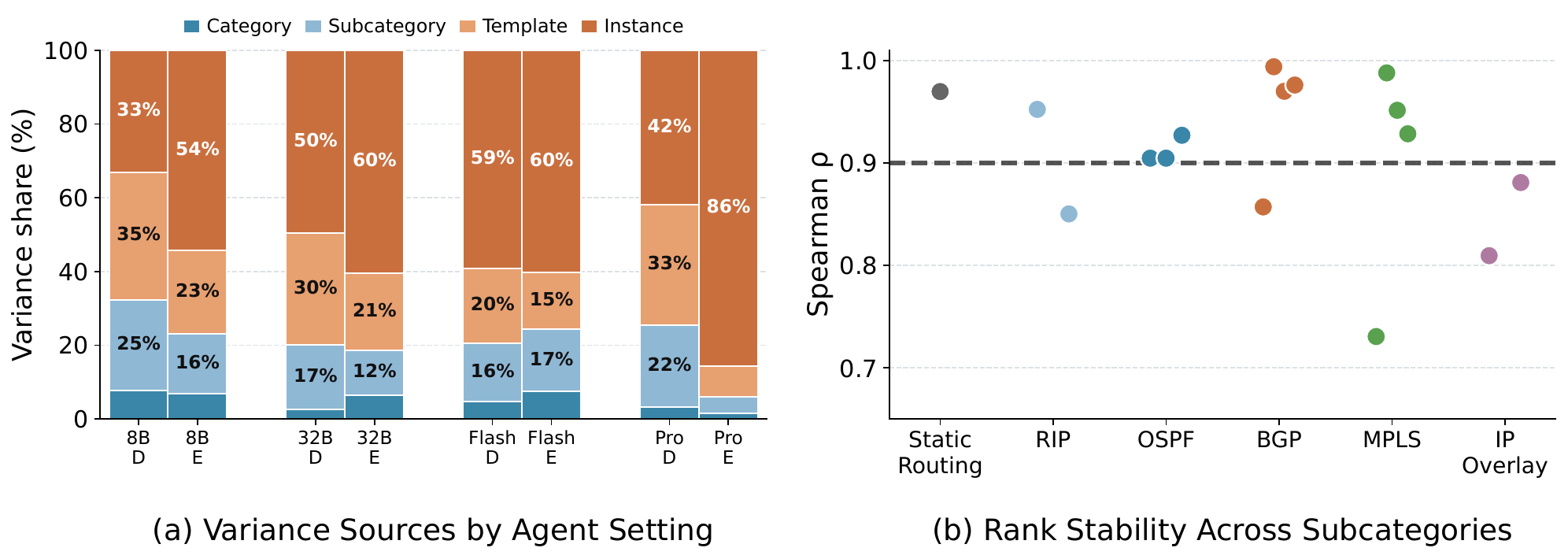}
  \caption{Generalizability analysis based on the test-case scores from the main evaluation. (a) Hierarchical score-variance decomposition; ``8B'' denotes qwen3-8B, ``32B'' denotes qwen3-32B, ``Flash'' denotes deepseek-v4-flash, and ``Pro'' denotes deepseek-v4-pro. The suffix ``D'' denotes reasoning disabled and ``E'' denotes reasoning enabled. (b) Spearman rank consistency for the 16 subcategories in the Table~V order; full subcategory names are omitted to keep the panel legible. The dashed line denotes $\rho=0.9$.}
  \label{fig:generalizability}
\end{figure}

Using the test-case scores from the main evaluation, we address this question with a variance decomposition and a cross-subcategory ranking analysis.
Figure~\ref{fig:generalizability} summarizes both analyses.

\paragraph{Variance decomposition across task-taxonomy levels}
Following the hierarchy in the task taxonomy, we decompose score variance across four levels: category, subcategory, template, and instance.
Fig.~\ref{fig:generalizability}(a) shows that the top-level category component accounts for 1.5\%--7.7\% of total variance, with a median of 5.7\%.
The subcategory and template components together account for a median of 37.5\%, indicating that finer-grained task requirements explain more variation than broad categories in this benchmark.
Because the benchmark contains only six top-level categories, the category-level estimate should be interpreted as a descriptive result rather than a precise population-level estimate.
At the instance level, randomized parameter instantiations of the same template can produce noticeably different outcomes.
Instance-level variation accounts for 33.0\%--85.7\% of total variance, with a median of 56.7\%.
Because parameter instantiation and model decoding both vary across trials, this analysis does not isolate their separate contributions.
Moreover, zero-variance templates often have all scores equal to 0 or 1, so the apparent stability may reflect a floor effect (all failures) or a ceiling effect (all successes).

\paragraph{Consistency of agent-setting performance orderings across task subcategories}
For each subcategory, we order the eight agent settings by their mean test-case score and compare this ordering with the overall ordering using Spearman's rank correlation.
Fig.~\ref{fig:generalizability}(b) shows the resulting correlations across the 16 subcategories.
The points follow the row order in Table~V, while full subcategory names are omitted to avoid overlapping labels.
The median Spearman correlation is $\rho=0.928$ (mean $0.912$), and 11 of the 16 subcategories achieve $\rho \geq 0.9$.
These correlations indicate that settings with stronger overall performance generally retain stronger relative performance within individual subcategories.

\begin{center}
\fcolorbox{gray!45!black}{gray!8}{\begin{minipage}{0.96\linewidth}
\small
\emph{Answer to RQ4:} Within NetConfArena, the observed correlations suggest stability of relative agent-setting rankings across the evaluated subcategories.
Randomized instances of the same template can yield different absolute outcomes.
\end{minipage}}
\end{center}

\end{document}